\documentclass[pra,twocolumn,showpacs]{revtex4}

\usepackage{graphicx}
\usepackage{amsmath}
\usepackage{amssymb}

\def\B#1{\left(#1\right)}

\begin{document}

\title{A quantum phase transition in a quantum external field: Superposing two magnetic phases}

\author{Marek M. Rams$^{1,2,3}$, Michael Zwolak$^4$, and Bogdan Damski$^1$}
\affiliation{$^1$\mbox{Los Alamos National Laboratory, Theoretical Division, MS B213, Los Alamos, New Mexico, 87545, USA}\\
$^2$\mbox{Vienna Center for Quantum Science and Technology, Faculty of Physics,
University of Vienna, Vienna, Austria}\\
$^3$\mbox{Institute of Physics, Jagiellonian University, Reymonta 4, 30-059 Krak\'ow,  Poland}\\
$^4$\mbox{Department of Physics, Oregon State University, Corvallis, OR 97331, USA}}

\begin{abstract}
We study an Ising chain undergoing a quantum phase transition in a {\it quantum} 
magnetic field. Such a field can be emulated by 
coupling the chain to a central spin initially in a superposition state.
We show that -- by adiabatically driving such a system -- one can prepare a quantum superposition of any
two ground states of the Ising chain. In particular, one can end up with the 
Ising chain in a superposition of ferromagnetic and paramagnetic 
phases -- a scenario with no analogue in prior studies of quantum phase
transitions. Remarkably, the resulting magnetization of the chain encodes the position of the critical point and universal critical exponents, as well as the ground state fidelity. 
\end{abstract}

\maketitle

Quantum phase transitions (QPTs) occur when dramatic changes in 
the ground state properties of a quantum system are induced by 
a tiny variation of an external parameter, such as the magnetic field in spin systems \cite{sachdev} 
or the intensity of a laser beam in cold atom simulators of Hubbard-like models \cite{LewensteinAdv}.  
In all current studies of QPTs, the external parameter is assumed to be classical, i.e., 
it has a well-defined instantaneous value. However, the field inducing a QPT
can be quantum as well, taking on different 
values by virtue of being in a superposition of states. In fact, tremendous progress with the preparation and 
manipulation of cold atom/ion systems will allow for creation of 
scenarios where the quantum nature of the ``external'' parameter will
play a {\it central} role. 

For instance, cavity-QED systems offer intriguing possibilities to study quantum control parameters 
\cite{Ritsch2005,Maciek2008,Esslinger2007}. 
In these systems, photons -- bouncing off two parallel mirrors -- interact with ultracold atoms.
If the number of photons in the cavity does not fluctuate, atoms experience an ``external'' 
periodic potential  $\cos^2({\bf kx})$, whose amplitude is 
proportional to the number of intra-cavity photons (${\bf k}$ is the photon
wave-vector).  Atoms in such a
system would be either in the superfluid phase or in the Mott insulator
phase \cite{LewensteinAdv}. It may be possible, however, to 
create a coherent superposition of the intra-cavity photonic states, giving rise to quantum 
fluctuations in the number of photons between the mirrors. The atoms would then be exposed 
to a coherent superposition of periodic potentials with the same period but differing amplitudes. 
In this case, one can have  atoms in a superposition of two quantum
phases, i.e., simultaneously in superfluid and Mott insulator ground states \cite{Ritsch2005}. 
Such a situation has no counterpart in traditional studies 
of QPTs where the system is either in one phase or another.  

An analogous phenomenon can be envisioned in central spin models.
These models are  used to describe 
qubit -- environment interactions in nitrogen-vacancy centers in diamond \cite{HansonScience},  
quantum dots in semiconductors \cite{CentralSpin1,CentralSpin2},  NMR experiments
\cite{JingfuPRL}, etc. 
The focus is typically on the  loss of coherence of the qubit 
while ignoring the environmental degrees of freedom. We will take the opposite perspective and 
explore the quantum state of the environment 
subjected to an effective 
quantum potential originating from the central spin. 
For an experimental study of such a 
scenario, one needs a well-controlled system, which we expect will be 
delivered in the foreseeable future by ion simulators of spin chains \cite{ion_sim}.

\begin{figure}[t]
\includegraphics[width=8cm]{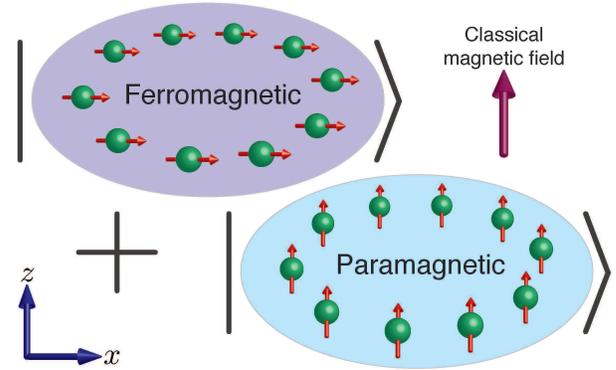}
\caption{Schematic of a superposition of two different quantum phases in an
Ising chain (in the paramagnetic phase spins try to align with the classical magnetic field, 
which is oriented in the $z$ direction here; in the ferromagnetic phase 
spin interactions try to align spins perpendicular to the field; 
see the Hamiltonian in Eq. (\ref{H_Ising}) for details).
One can prepare such a state by adiabatically evolving the chain in the presence of a 
central spin followed by a measurement of the spin.}
\label{fig1}
\end{figure}

\begin{figure}[t]
\includegraphics[width=\columnwidth]{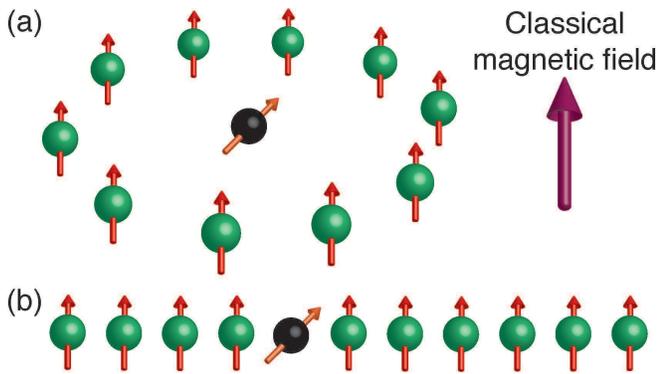}
\caption{(a) Schematic of the central spin model in a classical magnetic field: the central spin is equally 
coupled  to all the spins-1/2 arranged on  
a ring. (b) Possible realization of the central spin model in a linear  ion chain. 
The ions emulate the effective spins-1/2. 
The couplings between the effective spins-1/2 are optically engineered to be the same as in the (a) panel.
One of the ions is differently coupled to the rest of the chain to play the role of 
the effective central spin. The effective magnetic field is also optically engineered.
}
\label{fig2}
\end{figure}

\section{Results}
{\bf The model.}
We will discuss the most striking 
consequence of a QPT in a quantum potential: The possibility of having
the system in a superposition of ground states belonging to different phases, as shown in Fig.~\ref{fig1}. We 
consider a quantum Ising chain uniformly coupled to a (central) spin-1/2
(Fig. \ref{fig2}): 
\begin{equation}
\hat H = -\sum_{n=1}^N\left(\sigma^x_n\sigma^x_{n+1} + \hat g \sigma^z_n \right),
\label{H_Ising}
\end{equation}
where  $N\gg1$ is the number of spins arranged on a periodic ring. 
The central spin contribution  is contained in the effective magnetic field  
operator
\begin{equation}
\hat g = g+\delta\sigma^z_{S},
\label{ghat}
\end{equation}
where $g$ is the (classical) magnetic field strength and $\delta\sigma^z_S$
is the quantum component of the field generated by an Ising coupling to the central spin ($0<\delta\ll1$). 
Without the coupling to the central spin, i.e., when $\hat H(g,\delta=0) \equiv \hat H_I(g)$, the Ising chain in the ground
state is either in the ferromagnetic phase ($|g|<1$) or in the paramagnetic 
phase ($|g|>1$), with critical points at $g_c=\pm1$. 

The successful implementation of a recent proposal simulating arbitrarily-connected spin 
models in linear  ion chains \cite{ion_sim} will put the system we consider within experimental reach. 
It is interesting to note that there is no need 
to arrange ions, i.e., effective spins, on the ring to simulate the
Hamiltonian (\ref{H_Ising}), see  Fig. \ref{fig2}(a).
So far the simulation of an $N=6$ Ising chain with long-range
interactions between  the effective spins-1/2 has been demonstrated
\cite{MonroeNature}, analogous to that shown in Fig. \ref{fig2}(b).
The proposal is scalable and it is expected to allow for quantum 
simulation of models  with $N\gg1$ effective spins.

{\bf Preparation of the superposition state.}
QPTs can be studied either by diagonalizing the Hamiltonian for a fixed set of
coupling parameters or by adiabatically evolving the system from an
easy-to-prepare ground state, which is especially relevant in cold atom
experiments \cite{LewensteinAdv}. We take the latter approach, as the former
will always force the central spin to point in either $+z$ or $-z$ direction for any $g\neq0$
because $[\hat H,\sigma^z_S]=0$.

We assume that at $t=t_i$ the chain is prepared in a ground state and its coupling 
to the central spin is turned off, which provides freedom to engineer the state
of the central spin. The composite wave function is 
$|\psi(g(t_i))\rangle=|S\rangle |g(t_i)\rangle$, where $|g\rangle$ is a
ground state of $\hat H_I(g)$ and the central spin state is
$$
|S\rangle = c_\uparrow e^{i\phi_\uparrow\B{t_i}}\left|\uparrow\right\rangle + 
c_\downarrow e^{i\phi_\downarrow\B{t_i}}\left|\downarrow\right\rangle, \  c_\uparrow^2 + c_\downarrow^2=1,
$$
where $c_{\uparrow,\downarrow}>0$.
By  changing 
both the bias field $g$ and the coupling $\delta$, 
the wave-function evolves according to 
$$
|\psi(g(t))\rangle=\hat T \exp\left(-i\int_{t_i}^tdt\,\hat H[g(t),\delta(t)]\right)|\psi(t_i)\rangle,
$$
where $\hat T$ is the time-ordering operator.

As was shown in Ref. \cite{BD2009_Decoh}, $|\psi(g(t))\rangle$ can be simplified. 
Considering  adiabatic evolution, we obtain 
\begin{equation}
|\psi(g(t))\rangle = e^{i\phi_\uparrow\B{t}}c_\uparrow\left|\uparrow\right\rangle|g+\delta\rangle +  
e^{i\phi_\downarrow\B{t}}c_\downarrow\left|\downarrow\right\rangle|g-\delta\rangle.
\label{psii}
\end{equation}
We thus study finite, i.e., gapped, systems so that adiabatic evolution is possible by changing $g(t)$ and $\delta(t)$ slow 
enough. In the state (\ref{psii}), the chain 
experiences an average magnetic field 
$$
\langle \hat g\rangle = g + \delta (c_\uparrow^2 - c_\downarrow^2)$$ 
with fluctuations 
$$\sqrt{\langle\hat g^2\rangle -\langle\hat g\rangle^2} = 2\delta c_\uparrow c_\downarrow.$$ 
In particular, this shows that once the desired coupling $\delta$ is
adiabatically reached, fluctuations of the quantum potential are fixed.

The state (\ref{psii}) is already a Schr\"odinger's cat state, where the two ``macroscopically'' distinct 
possibilities are the ferromagnetic and paramagnetic phases, both of which are coupled to the auxiliary
two-level system.  
Since we are interested in the QPT of the Ising chain, we will ``trace
out'' the central spin by measuring its state. If we will
do the measurement in the $\{\left|\uparrow\right\rangle,\left|\downarrow\right\rangle\}$ basis, the superposition 
will be destroyed and the state of the Ising chain will be one of the ground states $|g\pm\delta\rangle$. 
Measurement in any other basis will result in a superposition of Ising ground states at different magnetic fields.

We  assume  that the 
measurement will be done in the eigenbasis of the $\sigma^x_S$ operator:
$$
|+\rangle = \frac{\left|\uparrow\right\rangle+\left|\downarrow\right\rangle}{\sqrt{2}},\quad
|-\rangle = \frac{\left|\uparrow\right\rangle-\left|\downarrow\right\rangle}{\sqrt{2}}.
$$
In this basis,  
\begin{align*}
|\psi(g(t))\rangle =
&|+\rangle 
\frac{c_\uparrow e^{i\phi_\uparrow}|g+\delta\rangle + c_\downarrow e^{i\phi_\downarrow}|g-\delta\rangle}{\sqrt{2}}+\\
 &|-\rangle
\frac{c_\uparrow e^{i\phi_\uparrow}|g+\delta\rangle - c_\downarrow e^{i\phi_\downarrow}|g-\delta\rangle}{\sqrt{2}},
\end{align*}
where  we write $\phi_{\downarrow,\uparrow}$ as shorthand for $\phi_{\downarrow,\uparrow}\B{t}$.
Therefore, the measurement of the central spin in the state $|\pm\rangle$ 
leaves the chain in the state 
\begin{equation}
\frac{c_\uparrow e^{i\phi_\uparrow}|g+\delta\rangle\pm c_\downarrow e^{i\phi_\downarrow}|g-\delta\rangle}{
     \sqrt{1\pm2c_\uparrow c_\downarrow\cos\B{\phi_\uparrow-\phi_\downarrow}{\cal F}(g,\delta)}},
\label{phi_pm}
\end{equation}
where  ${\cal F}(g,\delta)=\langle g-\delta|g+\delta\rangle$ is a ground state
fidelity \cite{GuReview}, or simply fidelity, whose crucial role in this problem will be carefully  discussed below. 
Without loss of generality, we define it in such a way that ${\cal F}(g,\delta)>0$.   

Now we comment on the measurement of the central spin. The state $|\pm\rangle$ will occur with probability 
\begin{equation*}
P_\pm\B{\phi_\uparrow-\phi_\downarrow}=\frac{1}{2}\pm c_\uparrow c_\downarrow\cos\B{\phi_\uparrow-\phi_\downarrow}{\cal F},
\end{equation*}
which depends on the relative phase between the ground states in the superposition
(\ref{psii}).
Since the point of the measurement is to prepare a well-defined superposition
state of the Ising chain, we will  describe properties of 
the Ising chain after finding the central spin in, e.g.,  the $|+\rangle$ state.
Then, the Ising chain will be  in the state 
\begin{equation}
|{\rm Ising}\rangle =
\frac{c_\uparrow e^{i\phi_\uparrow}|g+\delta\rangle+c_\downarrow e^{i\phi_\downarrow}|g-\delta\rangle}{
            \sqrt{1+2c_\uparrow c_\downarrow\cos\B{\phi_\uparrow-\phi_\downarrow}{\cal F}(g,\delta)}}.
\label{phi_pm1}
\end{equation}
This  state  is the desired superposition of ferromagnetic
 and paramagnetic ground states when 
$$
g-\delta< g_c < g+\delta,
$$ 
which is depicted in Fig. \ref{fig1}. We propose to call such a state a
Schr\"odinger magnet. 
The possibility to create such a novel state of matter is offered by 
the quantum  magnetic field in Eq. (\ref{ghat}). Indeed, if there would
be no quantum component in the magnetic field, 
the wave function of the Ising chain after the adiabatic evolution  would correspond to either a
ferromagnetic or a paramagnetic phase ground state, but never to a
superposition of both.

{\bf Ising chain in the superposition state.}
For simplicity, we assume that the measurements on the Ising chain are performed immediately 
after measuring the central spin. 
The expectation value of an operator $\hat O$ in the state  (\ref{phi_pm1}) is 
\begin{equation}
O=\langle {\rm Ising}|\hat O|{\rm Ising}\rangle=\frac{O^s +2c_\uparrow c_\downarrow
\cos\B\Delta O^{+-} }{1+2c_\uparrow c_\downarrow\cos\B\Delta\mathcal{F}},
\label{meanO1}
\end{equation}
where $\Delta=\phi_\uparrow-\phi_\downarrow$ and 
\begin{align*}
&O^s = c_\uparrow^2\,O^{++} + c_\downarrow^2\,O^{--}, \\
&O^{\pm \pm} =\langle g\pm \delta |\hat O |g\pm \delta \rangle,
\end{align*}
is the 
``standard'' average, and 
$$O^{+-}=\langle g+\delta |\hat O |g-\delta \rangle$$ 
designate the cross  term that arises. 
For clarity of presentation, we 
restrict ourselves to real $O^{+-}$ in Eq. (\ref{meanO1}), because $O^{+-}$ is always real for the
operators $\hat O=\sigma^z_n,\ \sigma^x_n,\ \sigma^x_n\sigma^x_{n+1}$ that we study. It
is not real, however, for all operators (e.g., for $\hat
O=\sigma^y_n$), and it is a straightforward exercise to extend our calculations 
to these cases.  While the standard average does not yield any new
information, the cross term provides a non-trivial correction absent in a
quantum phase transition in a classical field.

To further simplify the discussion, we average Eq. (\ref{meanO1}) over 
several  realizations where the appearance of the relative phase $\Delta$ 
of the superposition (\ref{psii}) is given by some probability distribution
$p\B{\Delta}$. For example, such averaging may appear due to preparation of the
central spin with random initial phases $\phi_{\uparrow,\downarrow}\B{t_i}$.
We assume below for simplicity that $p\B{\Delta}=\frac{1}{2\pi}$ for
$\Delta\in[0,2\pi)$.
We denote the result of such averaging as $\overline{O}$, and define its
variance through ${\rm var}(O)=\overline{O^2}-\overline{O}^2$. 
Finally, we introduce the notation 
$$
 O^{+-}_{\cal F} = O^{+-}/{\cal F},
$$
because for the $\hat O$ operators that we consider, 
$O^{+-}_{\cal F}$ is a well-defined non-zero  quantity in the thermodynamic limit in which 
fidelity typically tends to zero (see the Discussion section).

The phase-averaged observable and its variance are
\begin{equation*}
\begin{aligned}
&\overline{O}= \frac{\int_0^{2\pi} d\Delta\,p\B{\Delta} P_+\B{\Delta} O}{\int_0^{2\pi} d\Delta\,p\B{\Delta} P_+\B{\Delta}} =  O^s,\\
&{\rm var}(O)= \frac{\int_0^{2\pi} d\Delta\,p\B{\Delta} P_+\B{\Delta} O^2}{\int_0^{2\pi} d\Delta\,p\B{\Delta} P_+\B{\Delta}} -{\B{O^s}}^2 \\
      &\hspace*{29.5pt}=\left(O^s- O^{+-}_{\cal F}\right)^2 \left(\frac{1}{\sqrt{1-x^2}}-1\right),
\end{aligned}
\end{equation*}
with $x=2c_\uparrow c_\downarrow{\cal F}$. By expanding 
\begin{equation*}
\frac{1}{\sqrt{1-x^2}}-1 \approx \frac{x^2}{2},
\end{equation*}
we see that the square root of variance is proportional to fidelity when
$x\ll1$. The role of fidelity in our problem is discussed in the Discussion
section.
In the following, we use the exact solution of the Ising model to study expectation 
values of different observables in the superposition state  (\ref{phi_pm1}), 
see the Methods section  for details. 

We start by looking at $\hat O = \hat M_z = \sigma^z_n$. $M_z^{\pm\pm}$ terms have been
calculated in Ref. \cite{Pfeuty}
\begin{equation*}
M_z^{\pm\pm}=
\frac{1 + g\pm \delta}{\pi (g\pm \delta)}E(\chi_\pm)+ \frac{-1 + g\pm \delta}{\pi (g\pm \delta)}
K(\chi_\pm),
\end{equation*}
where $\chi_\pm=4 (g\pm \delta)/(1 + g\pm \delta)^2$, and 
$K$ and $E$ are elliptic functions of the first and the second kind, respectively. 
Above a large $N$ limit is assumed to simplify the expressions (see the Methods section for
the exact finite $N$  expressions).

\begin{figure}[t]
\includegraphics{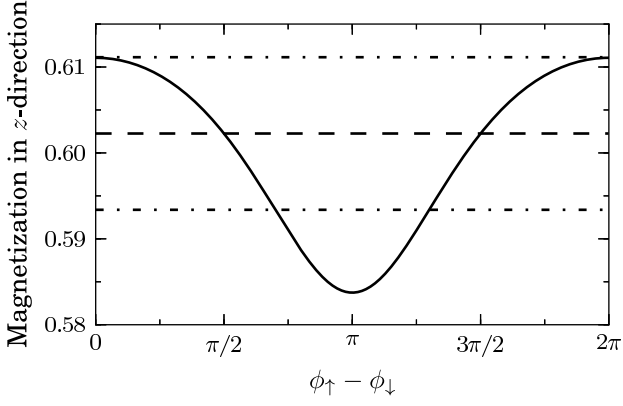}
\caption{Mean magnetization along the z-direction 
in a quantum superposition of different phases, $\hat O=\hat M_z =\sigma^z_n$. 
The plot shows exact results obtained from expressions 
listed in the Methods section.
The solid line shows $M_z$ evaluated  from  Eq. (\ref{meanO1}), the dashed line shows $\overline{M_z}=M_z^s$, and the
dashed-dotted line shows $\overline{M_z}\pm\sqrt{{\rm var}\B{M_z}}$.
We assumed $N=100$, $\delta=0.05$, $c_\uparrow= 1/2$, $c_\downarrow= \sqrt{3/4}$, and $g=1$.
For these parameters 
${\cal F}\approx0.41$, $\overline{M_z}\approx6.0\times10^{-1}$, and $\sqrt{{\rm var}(M_z)}\approx8.9\times10^{-3}$. 
}
\label{fig3}
\end{figure}

The cross terms can be obtained from the eigenequation
\begin{equation}
\hat H_I(g\pm\delta)|g\pm\delta\rangle=N\varepsilon(g\pm\delta)|g\pm\delta\rangle,
\label{HI}
\end{equation}
where $\varepsilon$ is the ground state energy per site.
Indeed, one gets from it
\begin{equation*}
M^{+-}_z = {\cal F}(g,\delta)\,\frac{\varepsilon( g-\delta)-\varepsilon( g+\delta)}{2\delta}.
\end{equation*}
In the limit of $N \to \infty$,  
$\varepsilon(g\pm\delta) = -\frac{2}{\pi} |1+g\pm\delta| E(\chi_\pm)$ \cite{Pfeuty}.
Consequently,
\begin{equation*}
M^{+-}_{z \cal F} = \frac{|1+g+\delta|}{\pi\delta}E(\chi_+)-\frac{|1+g-\delta|}{\pi\delta}E(\chi_-).
\end{equation*}
The dependence of magnetization on the relative phase of the superposition
and the variance of magnetization at $g=1$ are depicted in Fig. \ref{fig3}.

We mention in passing that similar expressions can be obtained for $\hat O= \hat C_x=\sigma^x_n\sigma^x_{n+1}$.
Indeed, it is known from Ref. \cite{Pfeuty} that 
\begin{equation*}
C_x^{\pm\pm}=
\frac{1 + g\pm \delta}{\pi}E(\chi_\pm)+ \frac{1 - g\mp \delta}{\pi}K(\chi_\pm),
\end{equation*}
and one can use again Eq. (\ref{HI}) to derive
$$
C_x^{+-} = \frac{{\cal F}(g,\delta)}{2}
\left[\frac{g-\delta}{\delta}\varepsilon(g+\delta)-
      \frac{g+\delta}{\delta}\varepsilon(g-\delta)\right].
$$
Since these results are analogous in structure to the ones already discussed, 
will not analyze  them. 

Next,  we  study spontaneous magnetization in the $x$-direction.
The system will acquire such a magnetization when a tiny field breaking the
$\sigma^x_n\to-\sigma^x_n$ 
symmetry of  the Hamiltonian is present. 
When necessary, 
we thus add a $-h\sum_{n=1}^N\sigma^x_n$ term to  $\hat H_I(g)$ and denote a ground state of the resulting Hamiltonian 
as $|g,h\rangle$. 
Without the quantum  magnetic field, $\delta=0$, the Ising chain acquires 
macroscopic  magnetization (along the direction of the symmetry breaking field $h$) 
only in the ferromagnetic phase. This magnetization
can also be  calculated by studying the correlation function  
\cite{Pfeuty} 
$$
\lim_{R\to\infty}\sqrt{|\langle g|\sigma^x_1  \sigma^x_R|g\rangle|}=
\lim_{h\to 0^+}\langle g,h|\sigma^x_n|g,h\rangle=
(1-g^2)^{1/8}.
$$ 
Importantly, it encodes the critical exponent $\beta=1/8$
\cite{ContinentinoBook}. 

\begin{figure}[t]
\includegraphics{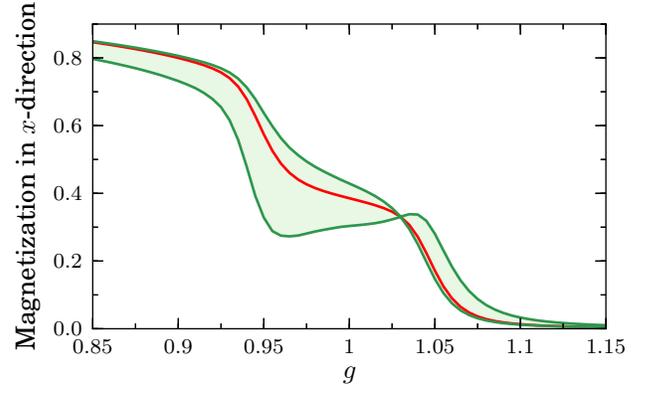}
\caption{Spontaneous magnetization in the $x$-direction in a quantum superposition of different phases.
The red dashed line is the ``standard'' average $M_x^s=c_\uparrow^2\,M^{++}_x+c_\downarrow^2\,M_x^{--}$.
The shaded area between the solid green lines marks the range of variation of $M_x$ 
due to variation of the relative phase $\phi_\uparrow-\phi_\downarrow$
(similar variation, but at a single magnetic field $g$,  is depicted in Fig. \ref{fig3}). 
This is a numerical result obtained with the ``periodic'' TEBD algorithm 
for $\delta=0.05$, $h=0.0001$, $c_\uparrow=c_\downarrow=1/\sqrt{2}$, 
 $N=100$, and $\chi=50$ (the cut-off parameter of the algorithm).
The spontaneous magnetization does not disappear for $g>1+\delta$,
 when both states in the superposition are in the paramagnetic phase, due
to the non-zero symmetry-breaking field $h$
(see Fig. \ref{fig5} for the $h\to0^+$ and $N\gg1$ limits; see the Methods
section for details).
}
\label{fig4}
\end{figure}

To study spontaneous magnetization in the presence of the superposition of ground states, we find numerically the states $|g\pm\delta,h\rangle$ 
using a periodic version \cite{VidalPBC} of the TEBD algorithm \cite{VidalAll1,VidalAll2}. 
Then, we calculate 
$M_x^{\pm\pm}= \left.\langle g\pm\delta,h|\sigma^x_n|g\pm\delta,h\rangle\right|_{h\approx0}$ and  
$M_x^{+-}=\left.\langle
g+\delta,h|\sigma^x_n|g-\delta,h\rangle\right|_{h\approx0}$. 
Naturally, for large enough systems, the standard result is reproduced 
by numerics:  
$$
M_x^{\pm\pm}\simeq[1-(g\pm \delta)^2]^{1/8}
$$ 
for $|g\pm \delta|<1$ and zero otherwise.
The results of TEBD calculations are plotted in Fig. \ref{fig4}. 
The presence of the cross term magnetization, resulting from the superposition of 
two ground states in Eq. (\ref{phi_pm1}), leads to {\it sizable} deviations from the
``standard'' average. 

To analyze this deviation more efficiently in the thermodynamic limit, 
we study the
asymptotic behavior of the two-point correlation functions:
\begin{equation}
M^{+-}_{x \cal F}= 
\lim_{R\to\infty}\sqrt{|\langle\sigma^x_1 \sigma^x_{1+R}\rangle^{+-}_{\cal F}|},
\label{crossterm}
\end{equation}
where $\langle\cdots\rangle^{+-}_{\cal F} = \langle g+\delta|\cdots|g-\delta\rangle/{\cal F}$.
It can be done using the exact solution of the Ising model through fermionization, 
where we express the correlator as a determinant of a
$2R\times2R$ block Toeplitz matrix, which is then numerically evaluated (see the Methods section for details). 

\begin{figure}[t]
\includegraphics{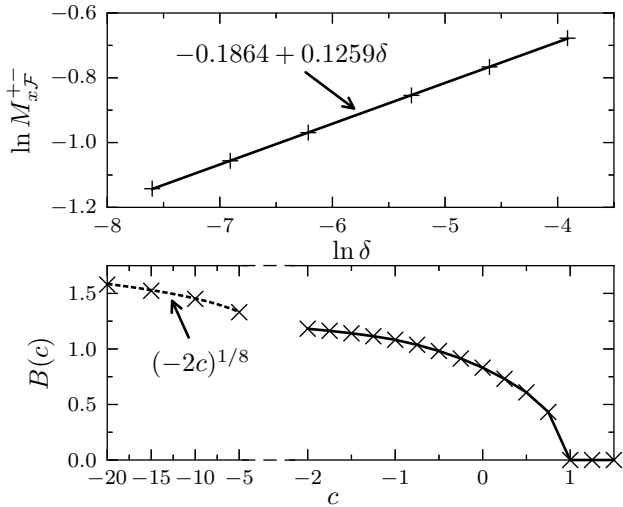}
\caption{Scaling properties of the spontaneous magnetization  cross term  in the
$x$-direction (\ref{Mxpm}).
Upper panel: Illustration that $M^{+-}_{x \cal F}(g,\delta)$ at the critical point  $g=g_c=1$  scales 
 as $\delta^{1/8}$.
Crosses show numerics based on Eq. (\ref{crossterm}), while the straight line is a fit. 
The same result is obtained near the critical point for other $c=(g-g_c)/|\delta|$ and $g<1+\delta$. 
Lower panel: Illustration of the scaling function $B(c)$ near the critical point and far 
away from it. Crosses show numerics, the solid line connects
them, and the dashed line is $(-2c)^{1/8}$. See the Methods section for details.
}
\label{fig5}
\end{figure}

As shown in Fig. \ref{fig5}, we find that the scaling of $M^{+-}_x$  around the critical point 
is consistent with the ansatz
\begin{equation}
M^{+-}_{x} = {\cal F}\delta^\beta B(c), \quad   c=\frac{g-g_c}{\delta},
\label{Mxpm}
\end{equation}
where $\beta=1/8$ and $g_c=1$ for the Ising chain that we study, and $B(c)$
is the scaling function.
It is nonzero when at least one of the superposed ground states 
is in the ferromagnetic phase, i.e., $B(c)\neq0$ for $c<1$. Far away from the
critical point, we observe that $M_x^{++}\approx M_x^{--} \approx M_{x{\cal F}}^{+-}$ and so 
 $B(c\ll-1) \simeq (-2c)^{1/8}$.   

\section{Discussion}
We have seen that the presence of the quantum external field allows for
creation of the superposition state of two distinct ground states in general
and two distinct phases in particular. If this
happens, expectation values are altered by the
cross terms. The magnitude of this effect can be sizable as depicted in 
Figs. \ref{fig3} and \ref{fig4}.

A fundamentally important question can now be answered:
what is the role of the system size in a quantum phase transition in a quantum field
and what critical information is imprinted onto the cross terms. 

To answer it, we note that all the cross terms that we studied 
are a product of the two terms: the ground state fidelity ${\cal F}(g,\delta)$ and a term that has
a well-defined non-zero value in the thermodynamic limit. Ground state fidelity, however, typically
disappears in the thermodynamic limit of $N\to\infty$ often invoked in the
context of quantum phase transitions. 

This is
known as Anderson orthogonality catastrophe after the seminal work reported in 
Ref. \cite{Anderson1967}. Therefore, we are interested in the studies of
systems for which $N\gg1$ (to see quantum criticality), but still $N<\infty$
(to avoid the catastrophe).
There are three options here, which we will discuss  below. Instead of providing 
specific results for fidelity of  the Ising chain, we provide general scaling results
to highlight the role of critical exponents in our problem and to keep the
discussion concise.

First, one can consider the limit of $\delta\to0$ taken at fixed $N\gg1$.
Then, fidelity reads \cite{ABQ2010,BarankovArXiv2009,PolkovnikovArXiv2010}
\begin{equation}
\ln{\cal F}(g,\delta) \sim - \delta^2 N^{2/d\nu}
\label{def_sus}
\end{equation}
near the critical point.
Here, $d$ is system dimensionality and $\nu$ is the critical exponent
(correlation length diverges as $|g-g_c|^{-\nu}$ near the critical point;
$d=\nu=1$ in the Ising chain that we consider).
Since fidelity is close to unity in this limit, the cross terms do not get
small.
One must remember, however, that they will be dominated by finite system
size corrections requiring a separate study, which is beyond the scope of this work.

Second, in the limit of $N\to\infty$ at fixed $\delta$ -- the one that we assumed in our calculations --
one can focus on the ``moderately'' large systems. To explain this term we
note that near the critical point (in the above-mentioned limit)
\cite{Fidel}
 \begin{equation}
\ln{\cal F}(g,\delta) \sim  -N\delta^{d\nu}.
\label{near_F}
\end{equation}
The crossover from Eq. (\ref{def_sus}) to Eq. (\ref{near_F}) happens near the
critical point when
\cite{Fidel}
$$
N\delta^{d\nu}\sim 1.
$$
We define the ``moderately'' large system to be just 
large enough to exhibit the scaling of fidelity with $\delta$ and $N$ given by
Eq. (\ref{near_F}) rather than Eq. (\ref{def_sus}). 
In the Ising case,  Eq. (\ref{near_F}) predicts $\ln{\cal F}\sim
-N\delta$ while Eq. (\ref{def_sus}) predicts $\ln{\cal F}\sim -N^2\delta^2$.
For  a ``moderately'' large
system fidelity shall not be too small
to erase the contribution of the cross term (see, e.g., Fig. \ref{fig4}). 

Third, one can study superpositions of two ground states from the same phase
far away from the critical point. There $\ln{\cal F}\sim
-N\delta^2|g-g_c|^{d\nu-2}$, and for $N\gg1$ fidelity can be kept close to
unity by a proper choice of $\delta$. The downside of this scenario  is that we loose the
possibility to superimpose two phases.

From the above discussion, we see that the critical exponent $\nu$ is
imprinted onto the cross term via fidelity. Also the critical exponent
$\beta$ is seen in the cross term $M_{x}^{+-}$ contributing to spontaneous magnetization.
The location of the critical point is most directly seen in the cross terms 
$M_z^{+-}$ and $C_x^{+-}$ through ``divergence'' of their second derivative over
$g$ taken at $g_c\pm\delta$. This is caused  by the singularity 
of the second derivative of the ground state energy per spin across the critical point.
This singularity will be rounded off in finite systems ($N<\infty$), but nevertheless there
shall be  pronounced peaks visible.
We also note that while the ``standard'' averages $M_z^{s}$ and $C_x^s$  also encode 
the position of the critical point, they do not encode the critical exponent $\nu$.

To observe the superposed phases, experiments will have to keep
decoherence to a minimum.
The decoherence rate of the state in Eq. (\ref{phi_pm1}) will depend on how well
the environment distinguishes
the two components, which will depend on the system size (see, e.g.,
Ref. \cite{DZZ}) and the overlap between
the two states (fidelity). Thus, $N$ cannot be too large and ${\cal F}$ cannot
be too small.
This is a similar issue to being able to observe the effect of the cross
term, which we discussed above. 
We thus do not expect decoherence to be overwhelming in a properly prepared
setup. Further, the system size
can be used as a parameter controlling decoherence, and its
manipulation should be sufficient to bring decoherence down to an
acceptable level. Looking from a different perspective, studies of 
decoherence of such a novel macroscopic quantum superposition are 
fundamentally interesting on its own, e.g., to boost understanding of the
quantum-to-classical transition. 
 
To conclude, we considered a quantum phase transition of an Ising
chain exposed to a quantum external field. This scenario can be used to create a new state of matter 
where the system is  simultaneously in two distinct quantum phases. Observables on the chain 
then take on forms that encode the ground state fidelity, the location of the critical
point, and the universal critical exponents of the system. These findings set the foundations 
for developing a scaling theory of quantum phase transitions in quantum 
fields. Recent advances in cold atom cavity-QED and ion traps may lead to experimental realization of 
superposed phases.

This work is supported by U.S. Department of Energy through the LANL/LDRD Program. 
MMR acknowledges support from the FWF SFB grant F4014.

\section{Methods}
Here we provide some technical details regarding our calculations.

The Ising Hamiltonian $\hat H_I(g)$ is diagonalized using the standard approach (see e.g. Ref. \cite{JacekPRL2005}). 
The Jordan-Wigner transformation,
\begin{equation*}
\sigma_n^z = 1-2 \hat c_n^\dagger \hat c_n,\quad 
\sigma^x_n = (\hat c_n + \hat c^\dagger_n) \prod_{m<n}(1-2 \hat c^\dagger_m \hat c_m),
\end{equation*}
where $\hat c_n$ are fermionic annihilation operators, transforms the Ising chain
to a free-fermion model. 
After applying the Fourier transform 
$$
\hat c_n = \frac{e^{-i\pi/4}}{\sqrt{N}}\sum_k \hat c_k e^{ikn},
$$
the  Hamiltonian takes the form:
$$
\begin{aligned}
\hat H_I(g) &= \sum_k (2 \hat c_k^\dagger \hat c_k -1) 
\left(g-\cos k \right) 
+ 
(\hat c_k^\dagger \hat c_{-k}^\dagger + \hat c_{-k} \hat c_{k})\sin k, \\
k &= \pm(2s + 1)\frac{\pi}{N},  \ \ s = 0,... ,\frac{N}{2} - 1. 
\end{aligned}
$$
Diagonalization of the Hamiltonian with the help of the Bogolubov 
transformation leads to the following  ground state wave-function
\begin{equation*}
\left| g \pm \delta \right> = 
\prod_{k>0} \left[ \cos \left(\theta_k^\pm / 2 \right) \left|0_k 0_{-k} \right> - \sin \left(\theta_k^\pm / 2 \right)\left |1_k 1_{-k} \right>  \right],
\end{equation*}
where $|m_k,m_{-k}\rangle$ describes the state with $m = 0,1$ pairs
of $c_k$ quasiparticles with momentum $k$ and 
\begin{equation*}
\tan\theta^\pm_k = \frac{\sin k }{g \pm \delta - \cos k }.
\end{equation*}

To prepare Fig. \ref{fig3}, we fix the system size $N$ and use the 
following exact expressions for  magnetization and fidelity
\begin{align*}
M^{\pm\pm}_z &= \langle g\pm\delta|\sigma^z_n|g\pm\delta\rangle = 
\frac{1}{N}\sum_k \cos\theta^\pm_k,\\
M^{+-}_z &= \langle g+\delta|\sigma^z_n|g-\delta\rangle = 
\frac{\cal F}{N}\sum_k \frac{  \cos\frac{\theta^+_k+\theta^-_k}{2}}{\cos\frac{\theta^+_k-\theta^-_k}{2}},\\
{\cal F}~&=~\langle g+\delta| g-\delta\rangle  ~=~\prod_{k>0} \cos\left(\frac{\theta^+_k - \theta^-_k}{2}  \right) > 0.
\end{align*}

To prepare Fig.  \ref{fig4}, we select   $g$, $\delta$ and $h$, and calculate the 
ground states $|g\pm\delta,h\rangle$ of the Ising chain exposed to
transverse and longitudinal magnetic fields. This is done through 
imaginary time evolution performed with the periodic TEBD algorithm.
A global phase of the wave-functions is then chosen to make 
${\cal F}=\langle g-\delta,h|g+\delta,h\rangle$ 
positive.
We then directly calculate  $M_x^{\pm\pm}$ and $M_x^{+-}$ (both are positive). 
Putting these results
into Eq. (\ref{meanO1}), one can calculate the 
spontaneous magnetization in the $x$-direction  in the superposition state 
(\ref{phi_pm1}). 
The  result is still dependent on the relative phase
$\phi_\uparrow-\phi_\downarrow$. 
When this phase is either $0$ or $\pm\pi$, spontaneous magnetization at any 
fixed $g$, $\delta$, and $h$ reaches an extremum. These extremal values are 
depicted by solid green lines in Fig. \ref{fig4}.

To prepare Fig. \ref{fig5}, we calculate correlation function 
$$
C^{+-}_{xx}(R)_{\cal F}
= |\langle\sigma^x_1 \sigma^x_{1+R}\rangle^{+-}_{\cal F}|=
|\langle  \Pi_{i=1}^R \hat b_i\hat a_{i+1}\rangle^{+-}_{\cal F}|,
$$ 
where we introduce $\hat b_n = \hat c_n^\dagger - \hat c_n$ and  
$\hat a_n = \hat c_n^\dagger + \hat c_n$.  
We study it, because $M^{+-}_{x \cal F}= \lim_{R \to \infty} \sqrt{C_{xx}^{+-}(R)_{\cal F}}$.

The next step is to use Wick's theorem extended to such a cross-correlation \cite{Wickextension}. 
It can be used as long as the overlap $\cal F$ is nonzero, which is the case in our calculations.
Then extending the results of Ref. \cite{BarouchMcCoy1971}, we find that 
$C^{+-}_{xx}(R)_{\cal F}$ can be expressed as a Pfaffian of a $2R \times 2R$  antisymmetric matrix, 
which can be converted into a determinant:
\begin{equation*}
\begin{aligned}
&C^{+-}_{xx}(R)_{\cal F} = \sqrt{\det \left[ A_R \right]}, \\
& A_R = \left[
\begin{matrix}
\langle  \hat b_m \hat b_n \rangle^{+-}_{\cal F} & \langle \hat b_m \hat a_{n+1} \rangle^{+-}_{\cal F} \\ 
\langle  \hat a_{m+1} \hat b_n \rangle^{+-}_{\cal F} & \langle \hat a_{m+1} \hat a_{n+1} \rangle^{+-}_{\cal F}
\end{matrix} \right] _{m,n=1\ldots R},
\end{aligned}
\end{equation*}
where $A_R$ is a block Toeplitz matrix.   Apart from a few special cases, it is not known how to calculate such a determinant  
analytically \cite{Its2009}. Thus, we  use numerics with a large enough $R$  to obtain a well-converged result.
We employ a continuous (i.e. $N\rightarrow \infty$) approximation for the elements of 
the Toeplitz matrix 
\begin{align*}
\langle \hat a_m \hat a_n \rangle_{\cal F}^{+-} &=  \frac{-i}{2\pi}\int_{-\pi}^{\pi}  dk 
\tan\frac{\theta^+_k-\theta^-_k}{2}  \  e^{i k (m-n)}, \\
\langle \hat b_m \hat a_n\rangle_{\cal F}^{+-} &= \frac{-1}{2\pi}\int_{-\pi}^{\pi}  dk \  \frac{e^{-i (\theta^+_k+\theta^-_k) /2} }{ \cos\frac{\theta^+_k-\theta^-_k}{2}} \   e^{i k (m-n)}, 
\end{align*}
and $\langle \hat a_m \hat a_n \rangle_{\cal F}^{+-}=
\langle \hat b_m \hat b_n \rangle_{\cal F}^{+-}$,  $\langle \hat a_m \hat b_n \rangle_{\cal F}^{+-}=
-\langle \hat b_n \hat a_m \rangle_{\cal F}^{+-}$.
Regarding the parameter $R$, we mention that 
it has to be of the order of  $500$ ($2000$) for $g=0.995$ and $\delta=0.01$ 
($g=1.005$ and $\delta=0.01$) in order for the results to be converged to the $R\to\infty$ limit. 
For every $g$ and $\delta$ sufficiently large $R$ is chosen to calculate data for Fig. \ref{fig5}.

We also mention that  we verified the Pfaffian-based numerics with 
a direct numerical calculation using the TEBD algorithm. For systems composed 
of about $100$  spins, for which the TEBD algorithm can still be efficiently applied, 
spontaneous magnetization from both calculations agree.

Finally, we provide definition of the elliptic functions that we use in the Results section:
\begin{align*}
K(x) &=\int_0^{\pi/2} \frac{d\phi}{\sqrt{1-x \sin^2 \phi}}, \\ E(x)&=\int_0^{\pi/2} d\phi \sqrt{1-x \sin^2 \phi}.
\end{align*}


\begin{thebibliography}{28}
\expandafter\ifx\csname natexlab\endcsname\relax\def\natexlab#1{#1}\fi
\expandafter\ifx\csname bibnamefont\endcsname\relax
  \def\bibnamefont#1{#1}\fi
\expandafter\ifx\csname bibfnamefont\endcsname\relax
  \def\bibfnamefont#1{#1}\fi
\expandafter\ifx\csname citenamefont\endcsname\relax
  \def\citenamefont#1{#1}\fi
\expandafter\ifx\csname url\endcsname\relax
  \def\url#1{\texttt{#1}}\fi
\expandafter\ifx\csname urlprefix\endcsname\relax\def\urlprefix{URL }\fi
\providecommand{\bibinfo}[2]{#2}
\providecommand{\eprint}[2][]{\url{#2}}

\bibitem[{\citenamefont{Sachdev}(2011)}]{sachdev}
\bibinfo{author}{\bibfnamefont{S.}~\bibnamefont{Sachdev}},
  \emph{\bibinfo{title}{Quantum Phase Transitions}}
  (\bibinfo{publisher}{Cambridge University Press},
  \bibinfo{address}{Cambridge, U.K.}, \bibinfo{year}{2011}).

\bibitem[{Lew()}]{LewensteinAdv}
\bibinfo{note}{Lewenstein, M., Sanpera, A., Ahufinger, V., Damski, B., Sen(De),
  A. \& Sen, U. Ultracold atomic gases in optical lattices: mimicking condensed
  matter physics and beyond. {\it Adv. Phys.} {\bf 56}, 243 (2007).}

\bibitem[{Rit()}]{Ritsch2005}
\bibinfo{note}{Maschler, C. \& Ritsch, H. Cold atom dynamics in a quantum
  optical lattice potential. {\it Phys. Rev. Lett.} {\bf 95}, 260401 (2005).}

\bibitem[{Mac()}]{Maciek2008}
\bibinfo{note}{Larson, J., Fern\'andez-Vidal, S., Morigi, G. \& Lewenstein, M.
  Quantum stability of Mott-insulator states of ultracold atoms in optical
  resonators. {\it New J. Phys.} {\bf 10}, 045002 (2008).}

\bibitem[{Ess()}]{Esslinger2007}
\bibinfo{note}{Brennecke, F., Donner, T., Ritter, S., Bourdel, T., K\"ohl, M.
  \& Esslinger, T. Cavity QED with a Bose-Einstein condensate. {\it Nature}
  {\bf 450}, 268 (2007).}

\bibitem[{Han()}]{HansonScience}
\bibinfo{note}{Hanson, R., Dobrovitski, V. V., Feiguin, A. E., Gywat, O. \&
  Awschalom, D. D. Coherent dynamics of a single spin interacting with an
  adjustable spin bath. {\it Science} {\bf 320}, 352 (2008).}

\bibitem[{Cen({\natexlab{a}})}]{CentralSpin1}
\bibinfo{note}{Bluhm, H. {\it et al.} Dephasing time of GaAs electron-spin
  qubits coupled to a nuclear bath exceeding 200 $\mu$s. {\it Nature Physics}
  {\bf 7}, 109 (2011).}

\bibitem[{Cen({\natexlab{b}})}]{CentralSpin2}
\bibinfo{note}{Cywi\'nski, \L. Dephasing of electron spin qubits due to their
  interaction with nuclei in quantum dots. {\it Acta Phys. Pol. A} {\bf 119},
  576 (2011).}

\bibitem[{Jin()}]{JingfuPRL}
\bibinfo{note}{Zhang, J., Peng, X., Rajendran, N. \& Suter, D. Detection of
  quantum critical points by a probe qubit. {\it Phys. Rev. Lett.} {\bf 100},
  100501 (2008).}

\bibitem[{ion()}]{ion_sim}
\bibinfo{note}{Korenblit, S. {\it et al.} Quantum simulation of spin models on
  an arbitrary lattice with trapped ions. {\it e-print} arXiv:1201.0776
  (2012).}

\bibitem[{Mon()}]{MonroeNature}
\bibinfo{note}{Islam, R. {\it et al.} Onset of a quantum phase transition with
  a trapped ion quantum simulator. {\it Nature Communications} {\bf 2}, 377
  (2011).}

\bibitem[{BD2()}]{BD2009_Decoh}
\bibinfo{note}{Damski, B., Quan, H. T. \& Zurek, W. H. Critical dynamics of
  decoherence. {\it Phys. Rev. A} {\bf 83}, 062104 (2011).}

\bibitem[{GuR()}]{GuReview}
\bibinfo{note}{Gu, S.-J. Fidelity approach to quantum phase transitions. {\it
  Int. J. Mod. Phys. B} {\bf 24}, 4371 (2010).}

\bibitem[{Pfe()}]{Pfeuty}
\bibinfo{note}{Pfeuty, P. The one-dimensional Ising model with a transverse
  field. {\it Ann. Phys.} {\bf 57}, 79 (1970).}

\bibitem[{\citenamefont{Continentino}(2001)}]{ContinentinoBook}
\bibinfo{author}{\bibfnamefont{M.~A.} \bibnamefont{Continentino}},
  \emph{\bibinfo{title}{Quantum Scaling in Many-Body Systems}}
  (\bibinfo{publisher}{World Scientific Publishing},
  \bibinfo{address}{Singapore}, \bibinfo{year}{2001}).

\bibitem[{Vid({\natexlab{a}})}]{VidalPBC}
\bibinfo{note}{Danshita, I. \& Naidon, P. Bose-Hubbard ground state: Extended
  Bogoliubov and variational methods compared with time-evolving block
  decimation. {\it Phys. Rev. A} {\bf 79}, 043601 (2009).}

\bibitem[{Vid({\natexlab{b}})}]{VidalAll1}
\bibinfo{note}{Vidal, G. Efficient classical simulation of slightly entangled
  quantum computations. {\it Phys. Rev. Lett.} {\bf 91}, 147902 (2003).}

\bibitem[{Vid({\natexlab{c}})}]{VidalAll2}
\bibinfo{note}{Vidal, G. Efficient simulation of one-dimensional quantum
  many-body systems. {\it Phys. Rev. Lett.} {\bf 93}, 040502 (2004).}

\bibitem[{And()}]{Anderson1967}
\bibinfo{note}{Anderson, P. W. Infrared catastrophe in Fermi gases with local
  scattering potentials. {\it Phys. Rev. Lett.} {\bf 18}, 1049 (1967).}

\bibitem[{ABQ()}]{ABQ2010}
\bibinfo{note}{Albuquerque, A. F., Alet, F., Sire, C. \& Capponi, S. Quantum
  critical scaling of fidelity susceptibility. {\it Phys. Rev. B} {\bf 81},
  064418 (2010).}

\bibitem[{Bar({\natexlab{a}})}]{BarankovArXiv2009}
\bibinfo{note}{Barankov, R. A. Quench dynamics as a probe of quantum
  criticality. {\it e-print} arxiv:0910.0255 (2009).}

\bibitem[{Pol()}]{PolkovnikovArXiv2010}
\bibinfo{note}{Gritsev, V. \& Polkovnikov, A. Universal dynamics near quantum
  critical points. {\it e-print} arXiv:0910.3692 (2009).}

\bibitem[{Fid()}]{Fidel}
\bibinfo{note}{Rams, M. M. \& Damski, B. Quantum fidelity in the thermodynamic
  limit. {\it Phys. Rev. Lett.} {\bf 106}, 055701 (2011).}

\bibitem[{DZZ()}]{DZZ}
\bibinfo{note}{Dziarmaga, J., Zurek, W. H. \& Zwolak, M. Non-local quantum
  superpositions of topological defects. {\it Nature Physics} {\bf 8}, 49
  (2012).}

\bibitem[{Jac()}]{JacekPRL2005}
\bibinfo{note}{Dziarmaga, J. Dynamics of a quantum phase transition: Exact
  solution of the quantum Ising model. {\it Phys. Rev. Lett.} {\bf 95}, 245701
  (2005).}

\bibitem[{Wic()}]{Wickextension}
\bibinfo{note}{Balian, R. \& Brezin, E. Nonunitary Bogoliubov transformations
  and extension of Wick's theorem. {\it Nuovo Cimento} {\bf 64}, 37 (1969).}

\bibitem[{Bar({\natexlab{b}})}]{BarouchMcCoy1971}
\bibinfo{note}{Barouch, E. \& McCoy, B. M. Statistical mechanics of the XY
  model. II. Spin-correlation functions. {\it Phys. Rev. A} {\bf 3} 786
  (1971).}

\bibitem[{Its()}]{Its2009}
\bibinfo{note}{Its, A. R. \& Korepin, V. E. The Fisher-Hartwig formula and
  entanglement entropy. {\it J. Stat. Phys.} {\bf 137}, 1014 (2009).}

\end{thebibliography}

\end{document}